\documentclass{ws-procs961x669}            
\usepackage{ws-toc,ws-multind}
\usepackage{threeparttable}
\usepackage{booktabs} 

\begin{document}








\wstoc{ Theoretical Modelling of Gamma-Ray Burst 090510 }{J. Saji}

\title{Theoretical Modelling of Gamma-Ray Burst 090510 }

\author{Joseph Saji}
\index{author}{Joseph Saji} 
\index{author}{Agnieszka Janiuk} 

\address{Center for Theoretical Physics, Polish Academy of Sciences\\
Warsaw 02-668, Poland
E-mail: \email{jsaji@cft.edu.pl}}

\author{Agnieszka Janiuk}
\address{Center for Theoretical Physics, Polish Academy of Sciences\\
Warsaw 02-668, Poland\\}

\author{Maria Giovanna Dainotti}
\address{Division of Science, National Astronomical Observatory of Japan, 2-21-1 Osawa, Mitaka, Tokyo 181-8588, Japan\\
Department of Astronomical Sciences, The Graduate University for Advanced Studies, SOKENDAI, Shonankokusaimura, Hayama, Miura District, Kanagawa 240-0115, Japan\\
Space Science Institute, Boulder, CO 80301, USA\\
Nevada Center for Astrophysics, University of Nevada, Las Vegas, NV 89154, USA\\
}

\author{Shubham Bhardwaj}
\address{Division of Science, National Astronomical Observatory of Japan, 2-21-1 Osawa, Mitaka, Tokyo 181-8588, Japan\\
Department of Astronomical Sciences, The Graduate University for Advanced Studies, SOKENDAI, Shonankokusaimura, Hayama, Miura District, Kanagawa 240-0115, Japan\\
}

\author{Gerardo Urrutia}
\address{Center for Theoretical Physics, Polish Academy of Sciences\\
Warsaw 02-668, Poland\\}

\begin{abstract}
 Gamma-ray bursts detected at high energies provide valuable insights into the emission mechanisms behind these still puzzling enigmatic events. In this study, we focus on GRB 090510, which is an unusual short GRB exhibiting plateau emission observed by the Fermi-LAT. Using the general relativistic magnetohydrodynamic code (HARM), we aim to infer the key properties of this GRB, such as the jet opening angle, the energetics, the Lorentz Gamma factor, the jet structure and its variability, and the progenitor parameters of the compact binary system. We explored both the 2D and 3D models and estimated the variability timescales. Our findings show that the predicted jet opening angle is within $88\%$ of the observed upper limit from observations, and the energetics are in general agreement with observed values when accounting for the evolution of jet opening angle with redshift. This work establishes the foundation for ongoing exploration, which will further align the theoretical model simulations with observational data.
\end{abstract}


\bodymatter


\section{Introduction}\label{intro}
\label{sec:Introduction}
Gamma-ray Bursts (GRBs) \cite{2015PhR...561....1K} are the intense pulses of high-energy radiation emitted by hyper-relativistic jets within our line of sight. Short GRBs are a subclass of GRBs that have their progenitors identified to be mergers of compact objects like Neutron Star (NS) and Black Hole (BH), with prompt emission lasting for $<$ 2 seconds. These jets are collimated with a half-opening angle $\theta_{jet}$. The jets exhibit complex and variable structures, and our understanding of them remains incomplete. Estimating the jet parameters like opening angle, variability, and energetics poses challenges because it requires high-quality multi-wavelength observations, which are often hindered by the limited availability windows in the observations. Therefore, even after decades of GRB observations, significant gaps remain in our understanding of these events. This leads us to approach the problem differently, aiming to combine numerical simulations with already available multi-wavelength observations to complement each other and gain deeper insights into the energy mechanism for these energetic events.

In this study, we employ theoretical model of a GRB central engine and jet based on General Relativistic Magneto-Hydrodynamic (GRMHD) simulations. We simulate a short GRB environment consisting of the accreting disk around a Kerr black hole leading to the production of ultra-relativistic jets. The initial characteristics will be tuned to produce jet characteristics similar to the one observed in GRB 090510. 

In this paper, we present the results from a 3D numerical simulation that we conducted to explain the characteristics of GRB 090510, along with insights and inferences drawn from the supplementary 2D simulations. \\


\section{Observational analysis}
\label{sec:observational-analysis}

\begin{figure}[h]
\centering
\includegraphics[width=0.69\textwidth]{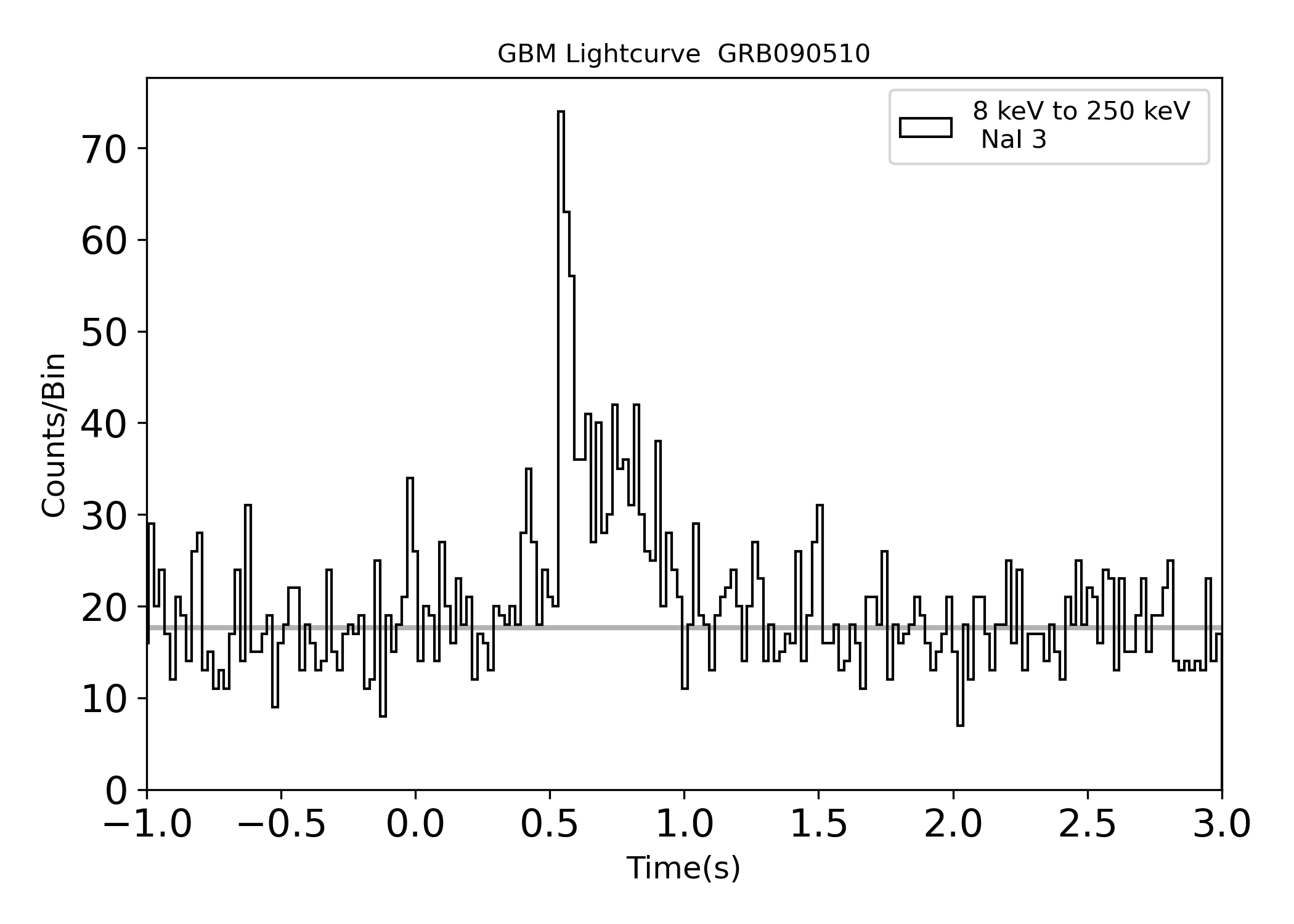}
\caption{Counts lightcurve of GRB090510 prompt emission from Fermi GBM observations}. 
\label{figure:fig1}
\end{figure}

GRB 090510 is a short GRB initially observed by the Fermi GRB space telescope. It has a $T_{90}$ of 0.6 s (Fermi GBM $T_{90}$). The measured redshift for this GRB is 0.903. 
The main reason for us to choose this GRB is that it is one of the short GRBs observed at high energies and it is also among the brightest short GRBs, making it a subject of significant scientific interest. We show in Fig. \ref{figure:fig1}, The prompt emission light curve (LC) of this GRB (8KeV - 250 KeV) produced from the Fermi GBM observations.


Since the luminosity computed by the numerical simulations is the bolometric luminosity, we need to calculate this by leveraging the multi-wavelength observations. For the total energy band, the luminosity is computed as:
\begin{equation}
L(E_{min},E_{max},t)= 4 \pi D_L^2(z) \, F (t) \cdot K(E_{min},E_{max}),
\label{eq: lx}
\end{equation}
where $D_L(z)$ is the luminosity distance computed in the flat $\Lambda{\rm{CDM}}$ cosmological model with $\Omega_M = 0.291$ and $h=0.70$ in units of $100$ $km$ $s^{-1}$ $Mpc^{-1}$, $F$ is the measured soft X-ray energy flux, $(E_{min}, E_{max})$ are the appropriate energy bandpass for each instrument we use. The \textit{K} is the \textit{K}-correction for the cosmic expansion \citep{Bloom2001}, defined below:

\begin{equation}
K=\frac{\int_{E_{min}/(1 + z)}^{E_{max}/(1 + z)}{\Phi (E)
dE}}{\int_{E_{min}}^{E_{max}}{\Phi (E) dE}},
\label{eq: kcorrection}
\end{equation}

\noindent

where $\Phi(E)$ is the energy spectrum of the afterglows, described by the Band function plus the power law. Thus, the bolometric luminosity, indicated with $L_{bol}$ is $L_{bol}$=$L_{XRT}+L_{GBM,LAT}$. We adopt the spectral fit parameters from the best-fit model identified by the combined Fermi GBM and LAT spectral analysis performed by the Fermi LAT/GBM collaboration \citep{Ackermann_2010_ed}. We find a bolometric luminosity of $4.5\times10^{53}$ erg $s^{-1}$ for GRB 090510. \\

The observational properties of GRB 090510, which we used to simulate the model, are summarized in Table \ref{tab:parameters}.

\begin{table}
\tbl{Summary of parameters for GRB 090510, comprising 
the time in which the 90\% of the total emission of the prompt is released, $T_{\rm{GBM,90}}$, start time ($T_{\rm{LAT,0}}$) and stop time ($T_{\rm{LAT,0}}$) of the LAT LC, rest-frame isotropic $E_{\rm{iso}}$, the redshift, and the corresponding distance in Gigaparsec.}
{\begin{tabular}{l@{\hspace{5cm}}l} \toprule
\multicolumn{2}{c}{090510} \\ 
\colrule
$T_{\rm{GBM,90}}$ (s) & 0.6 \\
$T_{\rm{LAT,0}}$ (s) & 1.0 \\
$T_{\rm{LAT,100}}$ (s) & 170.0 \\
$E_{\rm{iso}, LAT}$ (erg) & $5.46 \times 10^{52}$ \\
redshift ($z$) & 0.903 \\
distance (Gpc) & 5.86 \\
\botrule
\end{tabular}}
\label{tab:parameters}
\end{table}

\subsection{Jet Opening Angle From Observation}
\label{sec:jet-opening}

Opening angle $\theta_j$ is one of the main parameters of the jet we are focussing on in this work. We employ an analytical estimate of $\theta_{j}$ following Ref. \citenum{Pescalli2015MNRAS.447.1911P}, where the $E_{\rm{peak}} - E_{\rm{\gamma}}$ relation \citep{2004ApJ...616..331G} and the $E_{\rm{peak}} - E_{\rm{iso}}$ relation \citep{2002A&A...390...81A, 2009A&A...508..173A} have been used.

This gives us the relation for $\theta_{\rm jet}$ as,

\begin{equation}
  \cos\theta_{\rm jet} \approx  1 - \frac{5.078 \times 10^{11}}{E_{\rm{iso}}^{0.263}}.
\label{Eq:jet-angle}
\end{equation}

Thus, for our GRB, since we have the measure of the isotropic energy ($E_{\rm{iso}}$), $\theta_{\rm jet}$ can be calculated. This value of the jet opening angle needs to be corrected for the redshift evolution following Ref. \citenum{Lloyd2019MNRAS.488.5823L}. Using the Eq. \ref{Eq:jet-angle} and $E_{iso} = 4.5 \times 10^{53}$ (Section \ref{sec:observational-analysis}), we obtain a $\theta_{\rm{jet}}$ of $8.24^\circ {}^{+2.14}_{-1.73}$ corrected for redshift.

\subsection{Minimum Variability Timescale (MTS)}
\label{MTS-obser}
GRB LCs show a high degree of variability. The minimum timescale variability (MTS) refers to the shortest duration within which a significant change in the count rate occurs in the observed LC. The Bayesian Block (BB)\cite{Scargle_2013} method is employed in the LCs of the sample of GRBs to calculate MTS. BB divides LC into various time intervals of different time widths, closely following true underlying variation in the emission. The width of the least long-time bin is considered as MTS of the corresponding GRB. For our analysis, a count rate LC from the brightest NaI detector of Fermi GBM in the energy range of 10 KeV to 250 KeV is used. This LC is produced from the GRB data provided by The Fermi Science Support Center (FSSC) and processed using Fermitools. The minimum variability timescale of GRB 090510 has been calculated to be 32 milliseconds. \\


\section{Simulation Setup}
\label{sec:Simulation}

We perform GRMHD simulations using the code HARM\cite{Gammie_2003, Noble_2006, Sapountzis_2019} to model an accreting system around a rotating Kerr BH. The code follows the flow evolution by numerically solving the continuity, energy-momentum conservation, and induction equations in the GRMHD scheme. The code units are dimensionless, with $G$ = $c$ = $M$ = 1 for our simulations. The length scale in the code units is given by $r_g = GM/c^2$, and the time is given by $t_g = GM/c^3$, where $M$ is the mass of the BH.

\subsection{Model Parameters}
\label{sec:model parameters}
The accretion disk is modelled using the initial condition describing analytical Fishbone\& Moncrief (1976)\cite{1976ApJ...207..962F} solution, hereafter referred to as FM, which solves for a steady-state, pressure-supported fluid within the gravitational potential of a Kerr black hole. We use the polytropic equation of state \( P_g = K \rho^\Gamma\) in our model, where \( P_g \) is the gas pressure, \( \rho \) is the gas density. We use a value of \(\frac{4}{3}\) for \( \Gamma \) and $10^{-3}$ for $K$.
 The FM torus is embedded in a poloidal magnetic field and has a circular wire field configuration. The geometry of the imposed magnetic field has an important effect on the further evolution of the system. The non-vanishing component of vector potential is given by;

\begin{center}
\begin{equation}
    A_{\phi}(r, \theta) = A_0 \frac{(2-k^2)K(k^2) - 2E(k^2)}{k \sqrt{4Rr \sin \theta}} \label{eq-magfield}
\end{equation}

\[
    k = \sqrt{\frac{4Rr \sin \theta}{R^2 + r^2 + 2Rr \sin \theta}}
\]
\end{center}

$E$ and $K$ are complete elliptic functions and $A_0$ is the field normalisation constant. The radius of the circular wire is taken to be the $R_{max}$ of the FM torus. This field is normalized across the torus using a plasma-$\beta$ factor, which is the ratio between gas and magnetic pressure. The maximum $\beta$ value is fixed, and the magnetic fields are normalised to satisfy this criterion. In our model, the accretion disk feeds matter into the Kerr black hole as the angular momentum is transported outwards through magnetorotational instability (MRI). The accretion process subsequently drives the formation of a relativistic jet through the Blandford-Znajek \cite{1977MNRAS.179..433B} (BZ) mechanism.  The parameters of the model are provided in Table \ref{tab:Sim_Models}. 
The simulation was carried out in a 3D computational grid of $256\times128\times64$ in radial, polar, and azimuthal directions, respectively, with the outer boundary being set at 1000 $r_g$. 

Since the jets produced in our simulations are Poynting-dominated, we introduce two parameters: jet energetics parameter $\mu$ and jet magnetization parameter $\sigma$, which we will be using to quantify the properties of the simulated jet. 
The $\mu$ and $\sigma$ are defined as 

\begin{equation}
\mu = -\frac{T^r_t}{\rho u^r} \;\;\;\;\;\;\;\;  \sigma = \frac{(T_{\text{EM}})^r_t}{(T_{\text{gas}})^r_t}
\end{equation}
and are computed from the simulation given adequate components of the stress-energy tensor.

The quantity $\mu$ represents the total specific energy of the jet (the ratio of the total energy flux to the mass flux). Here, $T^r_t$ is the radial component of the energy-momentum tensor, representing the energy flux in the radial direction; $\rho$ is the rest-mass density and $U^r$ is the radial component of the four-velocity. 

The quantity $\sigma$ is the magnetization parameter, which gives the degree of magnetization in the jet. It is defined as the ratio of the electromagnetic energy flux to the gas energy flux. As for the luminosity of the jet, computed as the Blandford-Znajek luminosity, we calculate it at the given time snapshot as equal to $L_{BZ}=E_{unit} \int -T^{r}_{t} \sqrt{-g} d\theta d\phi$ and integrated over the BH horizon. $E_{\text{unit}} = M_{\text{unit}} \, c^{2} / t_{g}$, where $M_{\text{unit}} = \rho_s M_{\odot}$ is the mass unit determined by the density scaling factor $\rho_s$. For our physical models representing a general short GRB environment, we set $\rho_s = 1.5 \times 10^{-5}$.



\begin{table}
\tbl{Details of the 3D model used in this work to explain GRB 090510.}
{\begin{tabular}{cccccc} \toprule
Model & $r_{in}$ & $r_{max}$ & $\beta$ & $M_{disk}$ & $a$ \\ 
      & ($r_g$)    & ($r_g$)     &         & ($M_{\odot}$) &  \\ 

\colrule
\texttt{HD-0.10-B200-3D} & 12 & 18 &  200  & 0.1  & 0.95  \\
\botrule
\end{tabular}}
\label{tab:Sim_Models}
\end{table}

\subsection{Dynamical ejecta}
\label{sec:dynamical ejecta}
\begin{figure}
    \centering
    \includegraphics[width=0.69\linewidth]{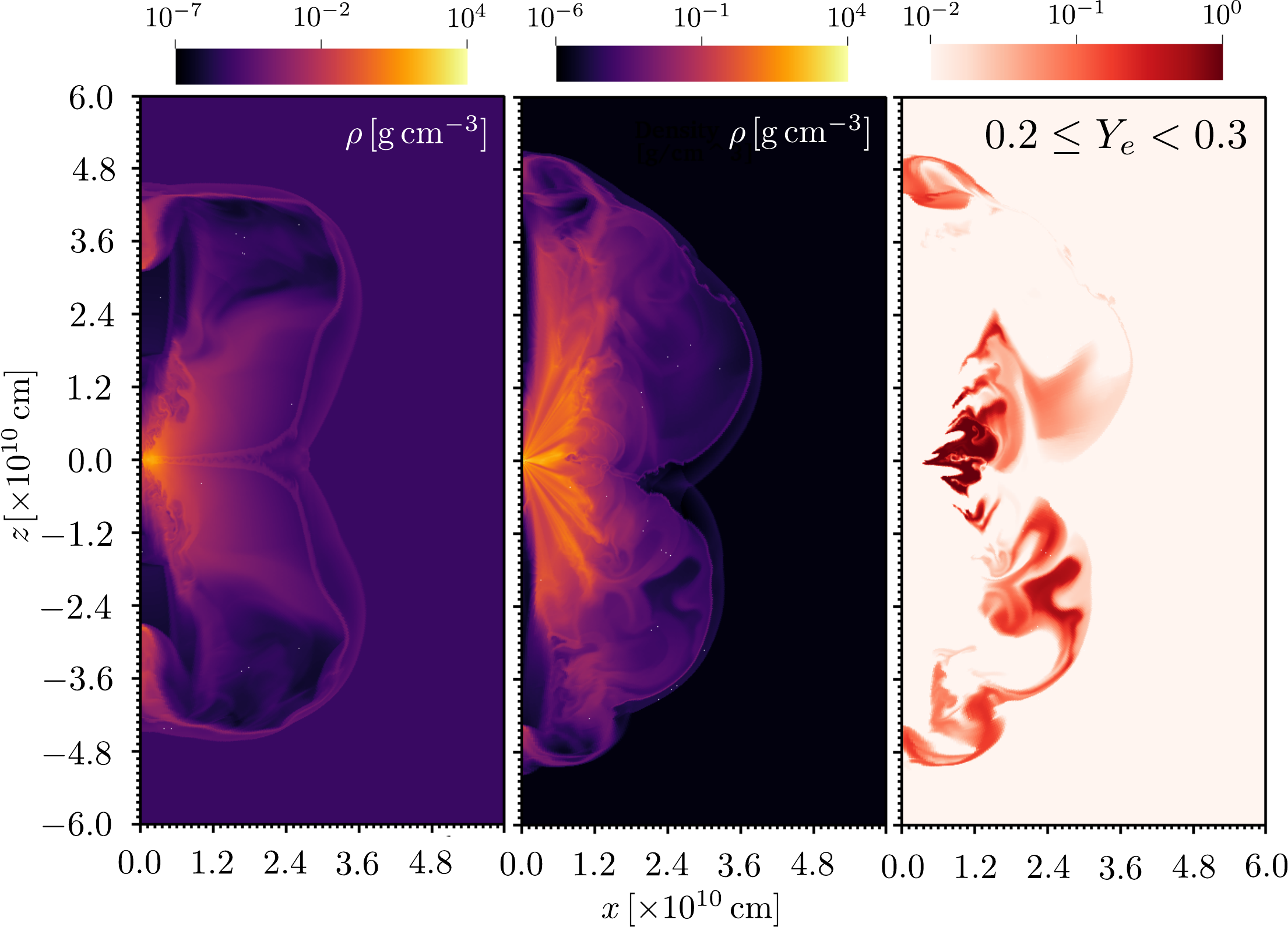}
    \caption{In the first panel on the left, the propagation of the jet inside a spherical homologously expanding ejecta is shown. In the central panel, we show the density map of the same jet but propagating alongside a post-merger disk wind. The third panel on the right shows the geometric distribution of lanthanide-rich outflows, or the red component of the kilonova, using passive scalars.}
    \label{fig:large-scale-ejecta}
\end{figure}

\begin{figure}
    \centering
    \includegraphics[width=0.85\linewidth]{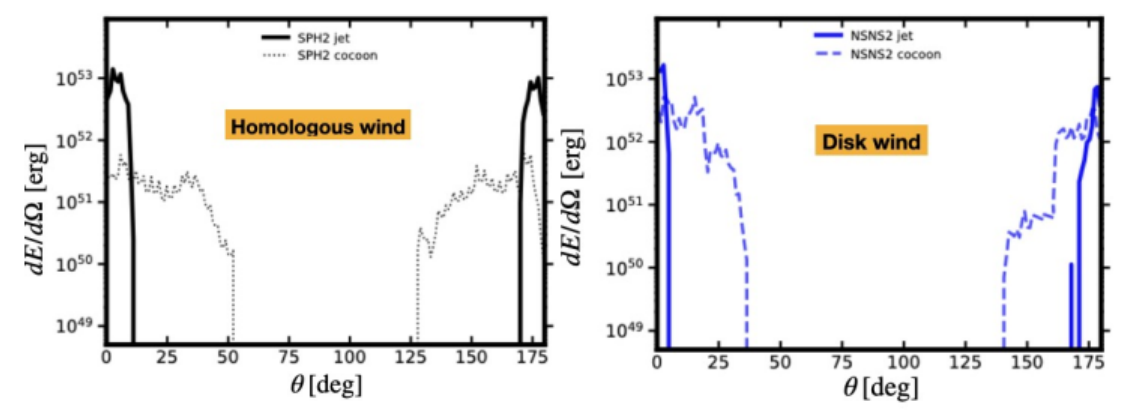}
    \caption{The comparison of the final structure of the jet propagating inside a homologously expanding ejecta and with the contribution of disk wind is considered. The main consequence of the disk wind model is the collimation of the jet, where the jet core ends up in a smaller angular region compared to the homologous case.}
    \label{fig:structure-large}
\end{figure}

The dynamics and structure of the jet are defined and affected by multiple parameters. Dynamic ejecta are a crucial feature that leads to further collimation of the jet. These ejecta are the material expelled during the inspiral phase of a binary neutron star merger. As the neutron stars spiral towards each other, their tidal interactions and the resulting shocks can eject a considerable amount of mass. This process occurs before the formation of a black hole. The ejected material, rich in neutrons, is a potential site for r-process nucleosynthesis, which can lead to the observable phenomena known as kilonova emissions. Understanding and estimating the mass and distribution of dynamic ejecta is challenging, especially in cases like GRB 090510, where direct kilonova observations are not available. However, we can model a simplistic application of this ejecta in the simulation environment using the approximations from theoretical models and merger simulations. Since we are dealing with an extremely narrow jet, the application of dynamic ejecta is a viable method to collimate our simulated jet. It is to be noted that we have not included the ejecta profile in the 3D model discussed here. However, we have run 2D tests with the models. We found out that this feature is essential to successfully collimate the jet. 
This is confirmed by recent simulations presented in Ref. \citenum{Urrutia2024}, which studied the effects of post-merger winds in comparison with pre-merger dynamical ejecta that we show in Figs~\ref{fig:large-scale-ejecta}-~\ref{fig:structure-large}.

In Fig~\ref{fig:large-scale-ejecta}, the first panel on the left illustrates the propagation of a jet with an initial opening angle of $\theta_j=15^\circ$. The jet is moving through a spherical ejecta with a mass loss rate of $\dot{M}=10^{-2}\,M_{\odot}$~s$^{-1}$, expanding homologously with a radial velocity $v=0.1\,c$. As the jet propagates, it and its cocoon spread laterally, while the density of the ejecta decreases with distance. The material is spread, and the structure of the jet covers a wide angular region (left panel of Fig~\ref{fig:structure-large}). In the central panel of Fig~\ref{fig:large-scale-ejecta}, we show a jet injected with the same initial opening angle as in the left panel, but here we consider the influence of a disk wind. This wind includes both radial and angular variations in density and velocity. Additionally, the pressure is affected by r-process nucleosynthesis. The post-merger environment is altered by this pressure. As the jet interacts with the surrounding medium, the pressure balance between the winds and the ejecta, along with the cocoon, is modified. This pressure equilibrium constrains the expansion of the jet, and finally squeeze it. In the right panel of Fig.~\ref{fig:large-scale-ejecta}, we show that the jet has become collimated. Specifically, the opening angle of the jet core, or the region of the jet, is narrower than the initial angle. The energy structure has been reshaped. These changes have important observational implications, for instance, in afterglows light curves, which are partly dependent on the jet opening angle and significantly affected by the modified jet structure, as estimated for example by 
Ref. \citenum{Urrutia2021ShortGRBS}. Finally, in the third panel on the right of Fig~\ref{fig:large-scale-ejecta}, we track the geometric distribution of lanthanide-rich outflows ($Y_e \simeq 0.25$) using passive scalars. This map qualitatively represents the red kilonova component, which has high opacity.\\

\hfill \\
\\


\section{Results}
\label{sec:Results}

Focusing on GRB 090510, the simulations were conducted over 44,000 $t_g$, equivalent to 0.6 seconds in real-time, aligning with the Fermi GBM's estimated $T_{90}$ for GRB 090510. A black hole mass of 3 $M_{\odot}$ was considered.

\begin{figure}[h]
\centering
\includegraphics[width=0.75\textwidth]{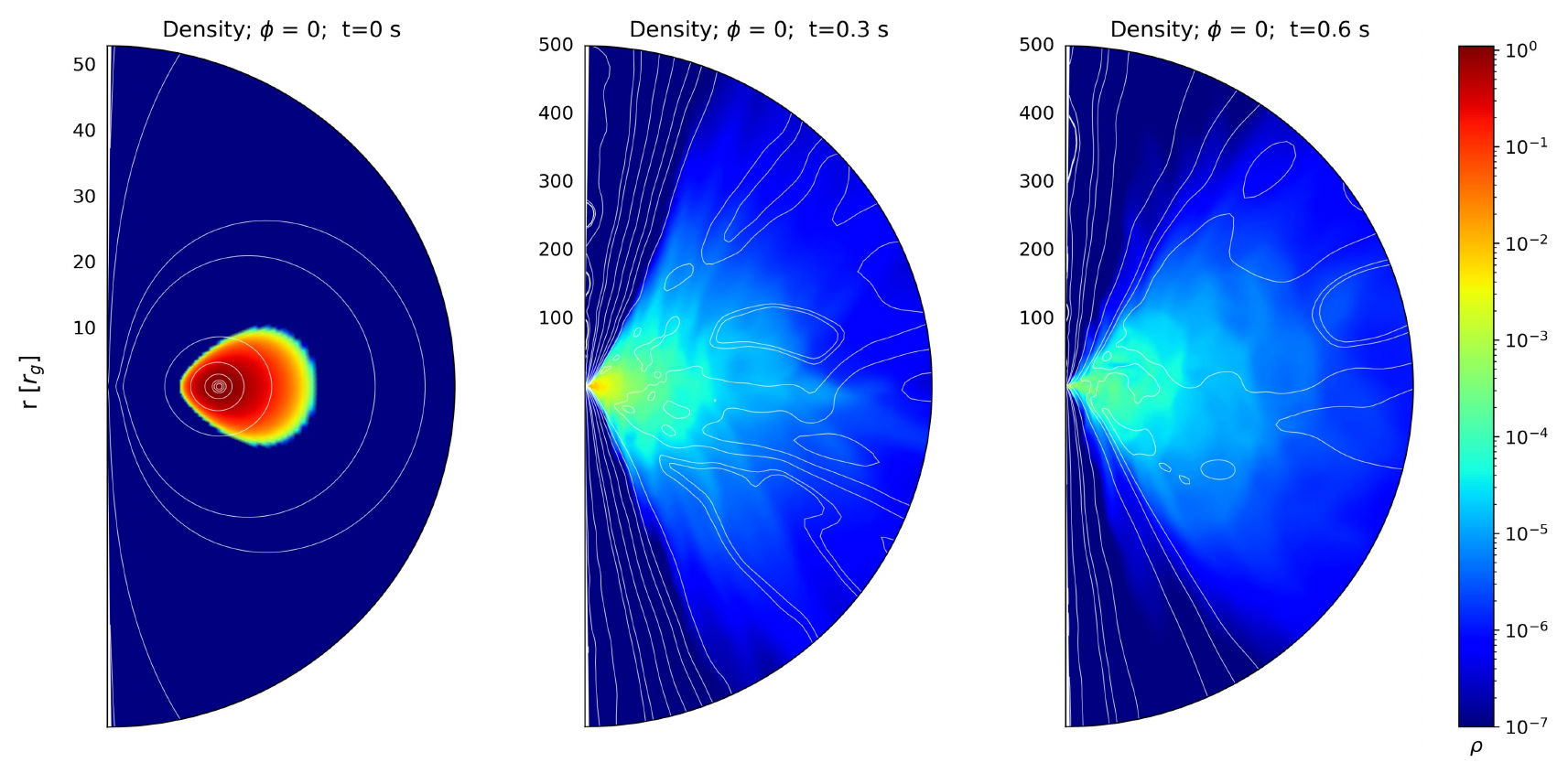}
\caption{Snapshots of density combined with magnetic flux contours for the 3D model HD-0.1-200-3D ($\phi = 0$) slice; The snapshots represent three different time instances: t=0s, t=0.3s, and t=0.6s. The density values are in code units}
\label{figure:Density_evol_3DModel}
\end{figure}

\begin{figure}[h]
\centering
\includegraphics[height=0.3\textheight]{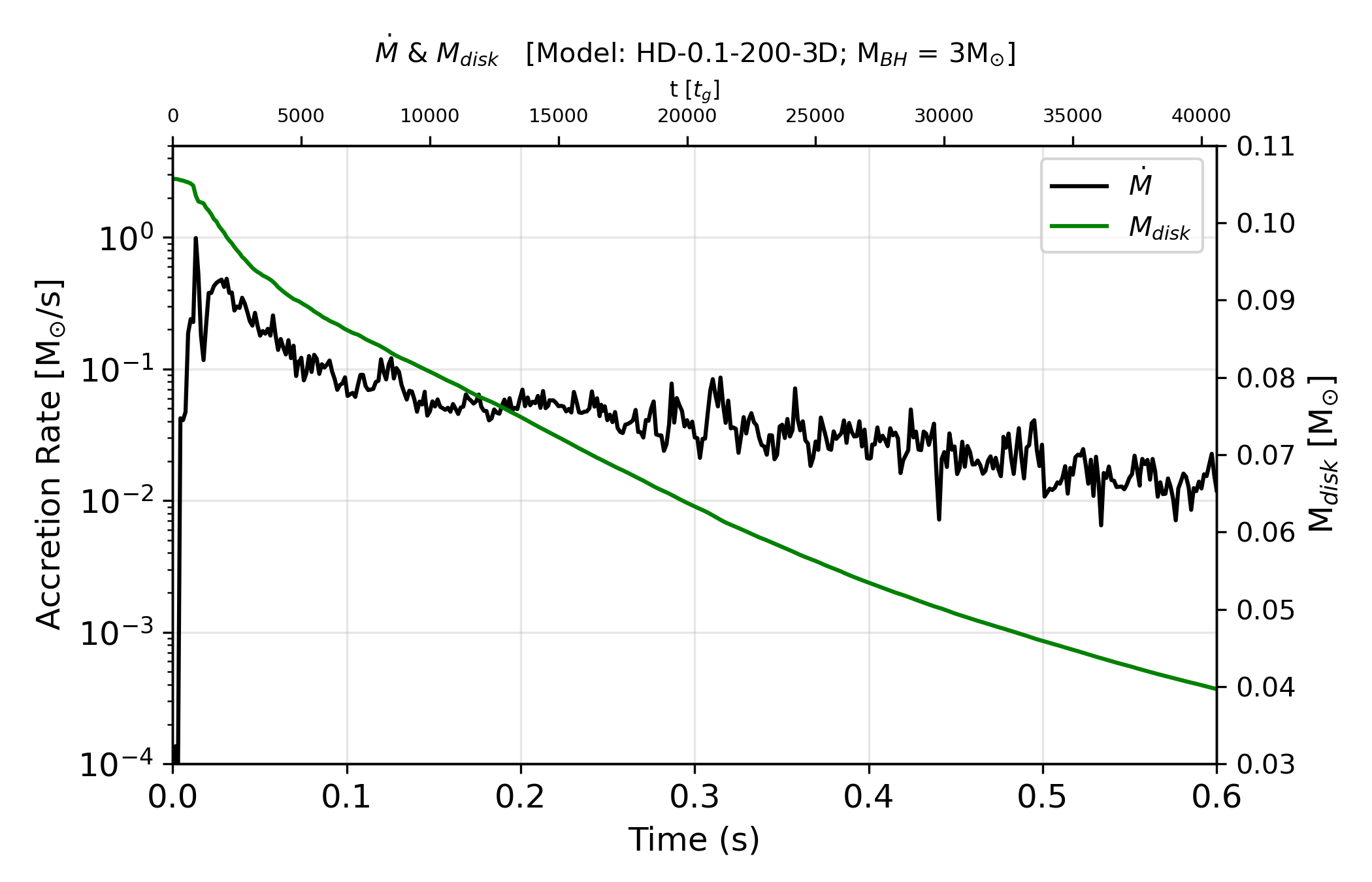}\\

\caption{Time evolution of Mass accretion rate ($\Dot{M}$) in black solid lines and disk mass ($M_{disk)}$) in green solid lines.} 
\label{figure:lumin}
\end{figure}

The primary focus of the discussion centres on the three-dimensional (3D) model \texttt{HD-0.10-B200-3D} without the dynamic ejecta. This model features an initial disk mass of $0.1\, M_{\odot}$ and a magnetic field scaled by the parameter $\beta$, normalized to 200. The evolution of density from the initial setup to the mid-point and final stage of the simulation for this model is illustrated in Figure~\ref{figure:Density_evol_3DModel}, corresponding to times of 0\,s, 0.3\,s, and 0.6\,s, respectively. Additionally, the time evolution of mass accretion rate and disk mass is depicted in Figure~\ref{figure:lumin}.

The initial disk mass of $0.1\,M_\odot$ was reduced to $0.04\,M_\odot$ at the end of $6\,\mathrm{s}$. The accretion rate is fairly stable, peaking initially at $1\,M_\odot\,\mathrm{s}^{-1}$ and slowly decreasing to the range of $10^{-2}\,M_\odot\,\mathrm{s}^{-1}$ towards the end of the simulation. The model produces a maximum $L_{Bz}$ of $3.7 \times 10^{53}$ erg/s. It is important to note that this luminosity differs from the observed isotropic luminosity $L_{iso}$. Since our simulations do not incorporate radiative transport mechanisms, it is challenging to determine the fraction of the total jet energy converted into radiation observable by telescopes. However, an analytical estimate can be derived from theoretical predictions regarding the radiative efficiency of Blandford-Znajek jets\cite{lloyd-ronning_constraints_2019} and the opening angle of the simulated jet. The resulting value is of the same order of magnitude and closely matches the $L_{iso}$ calculated for GRB 090510 from multi-wavelength observations, which is $4.5 \times 10^{53}$ erg/s.

\begin{figure}[h]
\centering
\includegraphics[height=0.3\textheight]{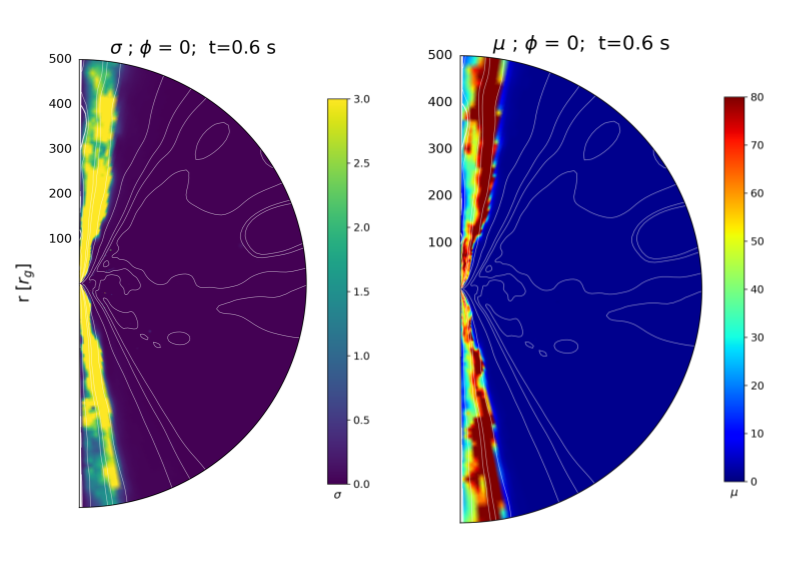}
\caption{Jet structure from the 3D model, quantified by $\mu$ and $\sigma$. Snapshot at  ($t=0.6s,\phi=0$) is shown, providing a detailed look into the jet structure towards the end time.}
\label{figure:mu_sigma}
\end{figure}

The black hole rotates with Kerr parameter a = 0.95. The spin value falls on the upper range of theoretical estimates for the stellar mass black holes but is required since GRB 090510 is an exceptionally bright short GRB. The BZ mechanism of jet production is powered by the extraction of rotational energy from the black hole, and thus, the high spin contributes to raising the luminosity of the jet to the desired range. The jet structure is studied using the jet energetics parameter $\mu$ and jet magnetisation parameter $\sigma$. The jets are produced as soon as the accretion begins and originate from the rotatory axis of the Kerr black hole. The jets here are pointing dominated, and their structure at the end time of the simulation is provided in Figure \ref{figure:mu_sigma}.

\begin{figure}[h]
\centering
\includegraphics[ height=0.3\textheight]{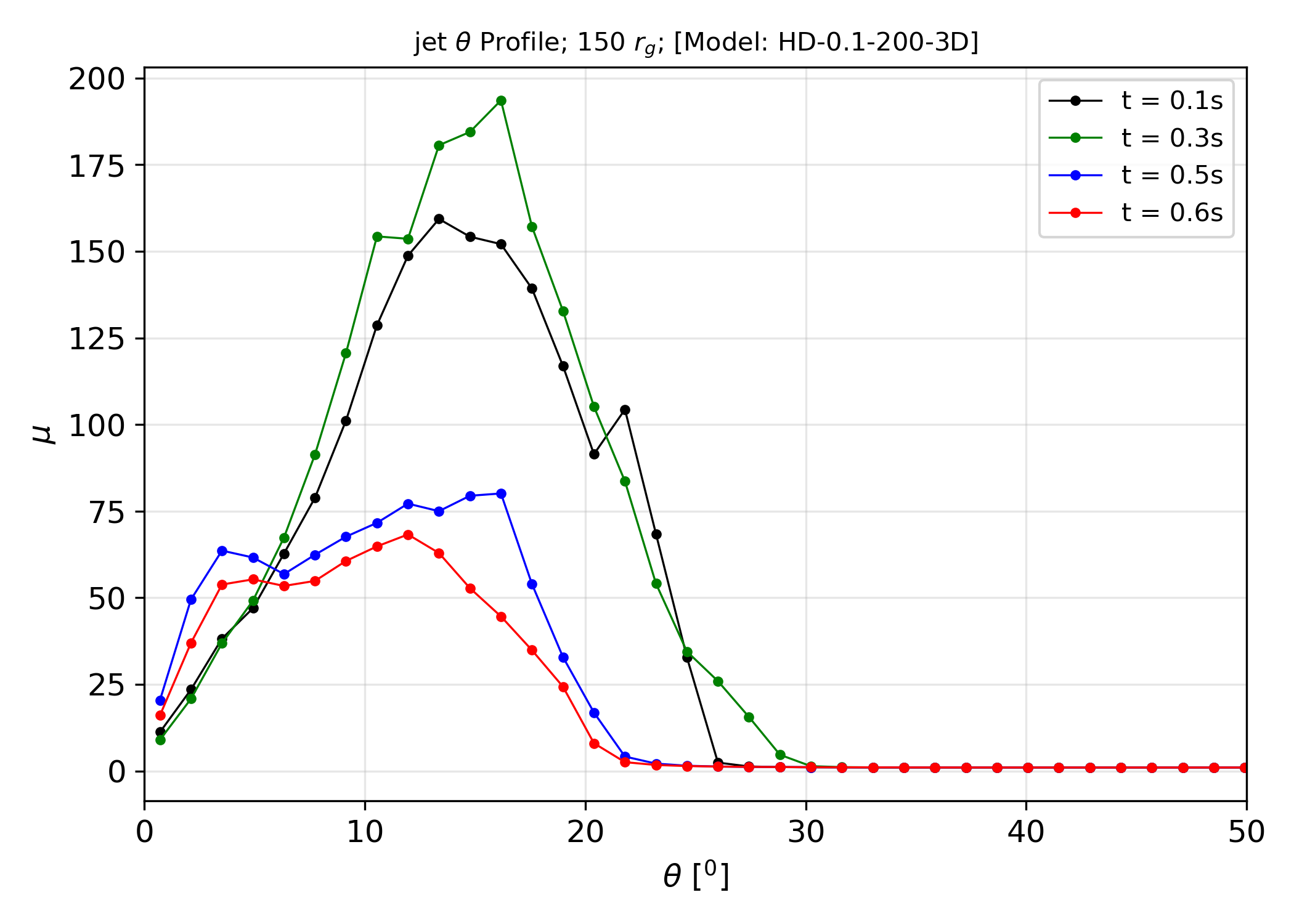}

\caption{Time evolution of $\mu$ (averaged over $\phi$) as a function of $\theta$. This plot illustrates the variations in jet structure at different times, highlighting the dynamic changes in $\theta_j$ as well as jet structure.}
\label{figure:mu_angle}
\end{figure}

Given that $\mu$ provides a direct estimate of the electromagnetic energy confined within the jet, we employ $\mu$ to delineate the jet region and determine the jet opening angle. The opening angle is defined as the angle $\theta$ that encompasses 75\% of the total jet energy. We have used this definition to calculate the opening angle for the model \texttt{HD-0.10-B200-3D}, and the results are provided in Table~\ref{lor_oA}. The jet stabilizes in the initial tens of milliseconds, and we calculate the opening angle by taking the time average of $\mu$ at different timescales and radii to perform a detailed estimate of the jet's opening angle (see Figure~\ref{figure:mu_angle}).  $\mu$ as a function of $\theta$ for four different times is plotted in Figure \ref{figure:mu_angle}. This gives a detailed picture of the evolution of jet cross-section over time. It is also to be noted that the jet has a hollow core ($2^\circ$ - $5^\circ$), which gets filled up over time and is almost non-existent by the end time of 0.6s. The jet becomes narrower over time and radius, and the opening angle calculated at the last tenth of a second (0.5-0.6\,s) is approximately $\sim 11.80^\circ$. This value is close but not exact to the opening angle estimate from the observations for GRB 090510. This led us to rely on other external interactions that the jet can have during its ejection process, which can lead to further collimation. One such option is the effect of dynamic ejecta, as described in Section \ref{sec:dynamical ejecta}.

 The jet energetics parameter $\mu$ is also employed to estimate the Lorentz factor ($\Gamma$) at infinity, assuming all energy is transformed to the baryon bulk kinetic \citep{Vlahakis_2003}. From the energy conservation along the field lines, we can conclude that the $\mu$ parameter is equivalent to the bulk Lorentz factor at infinity when all the energy is transformed into the kinetic energy in the jet. In order to calculate the Lorentz factor, we choose different fixed locations in the jet region (r,$\theta$) and take a time average of $\mu$. The results are provided along with $\theta_{jet}$ in Table~\ref{lor_oA}. \\

\begin{table}
\tbl{Luminosity($L_{Bz}$), Opening angle $\theta_{jet}$, Lorentz Factor ($\Gamma$) and Minimum timescale variability($MTS$) calculated for the 3D model \texttt{HD-0.10-B200-3D}. $\Gamma$ and $MTS$ are calculated at two different locations in the jet}
{\begin{tabular}{cccccc} \toprule
Model & $L_{BZ, Max}$ & $\theta{jet}$  & $\theta{jet}$   & $\Gamma$ & MTS \\ 
      & ($ergs/s$)    & (0.1-0.3s)     &   (0.5-0.6s)  &          & ($ms$) \\ 

\colrule
\texttt{HD-0.10-B200-3D} & $3.67\times10^{53}$ & $21.65^\circ$ &  $11.80^\circ$  & 70.95 {\tiny ($150r_g$, $8^\circ$)}  & 7.4 {\tiny ($150r_g$, $8^\circ$)} \\
&   &   &   & 66.40 {\tiny ($500r_g$, $8^\circ$)}  & 6.5 {\tiny ($150r_g$, $12^\circ$)} \\

\botrule
\end{tabular}}
\label{lor_oA}
\end{table}


Another important parameter that has the potential to directly correlate with observational results is the variability timescale. GRBs are inherently highly variable in nature, which is clearly manifested in their light curves. We have discussed the calculation of the Minimum Timescale variability (MTS) parameter in Section \ref{MTS-obser} and found the MTS of GRB 090510 to be $32\,\mathrm{ms}$. Identifying a simulation parameter that can estimate variability in a manner directly comparable to the observed MTS is challenging. Additionally, observed properties may be affected by significant noise and other external interactions that can obscure the true minimum variability timescale, potentially rendering it incompatible with quantities derived from simulations.

Nevertheless, we find that the jet energetics parameter $\mu$ remains a very suitable candidate for quantifying jet variability, as it directly measures the total energy available to the jet. Variations observed in the light curves arise from fluctuations in this energy and from collisions between different layers of material within the jet. Previous works (see Ref.~\citenum{James_2022, Janiuk_2021}) have utilized $\mu$ to quantify the MTS and compare it with observations. Following this methodology, we analyze the time evolution of $\mu$ at different locations within the jet and calculate the average duration of the peak widths of $\mu$ at their half-maximum. This average duration serves as a proxy for the jet's MTS. Using this method, we estimate an MTS of $6.5\,\mathrm{ms}$ from simulations, an order of magnitude lower than the observed MTS for GRB 090510. However, more sophisticated methodologies for MTS calculations can be considered. In this paper (see Ref .~\citenum{10.1093/mnras/stt241}), G. A. MacLachlan employed a novel wavelet analysis method, which appears to be more effective in cases with noisy light curves. Utilizing this wavelet analysis, the calculated MTS for GRB 090510 is $5\,\mathrm{ms}$, which is much closer to our simulation estimate. In summary, our \texttt{HD-0.10-B200-3D} model produces a stable jet lasting 0.6 seconds with a luminosity of approximately $10^{53}$ erg/s, closely matching the estimated values from observations of GRB 090510. The jet exhibits an opening angle of 11.80 degrees, which is within 88$\%$ of the observed upper limit for $\theta_{jet}$. Additionally, the minimum variability timescale in our simulation aligns with the observed timescale.


\section{Discussions and Conclusions}

Understanding jet properties, such as the jet opening angle, Lorentz factor, and variability timescales, is vital for uncovering the mechanisms behind GRB emissions. Reliance on direct observations for these properties is challenged by the limited number and scope of multi-wavelength observational campaigns. Simulations offer a complementary approach, enabling the exploration of jet dynamics beyond the constraints of observational data and facilitating the development of comprehensive models of GRB behaviour. By integrating simulations with observations, we can achieve a more nuanced understanding of GRB jets, thereby enhancing both the accuracy of jet characterizations and the predictive capabilities for unobserved bursts. In this paper, we have primarily discussed 3D simulations designed to model short GRB jets with properties comparable to GRB 090510, including energetics, variability timescale, and opening angle. The model HD-0.10-B200-3D, as detailed in Section~\ref{sec:Results}, was largely successful in replicating the aforementioned parameters, although it slightly overestimated the jet opening angle. Despite this limitation, the model remains a robust theoretical framework for explaining GRBs similar to GRB 090510. Collimating jets produced from GRMHD simulations to achieve narrower angles presents significant challenges. Observational analyses have determined opening angles as small as $1^\circ$ to $10^\circ$, which are difficult to attain in simulations due to inherent numerical and physical complexities. The incorporation of dynamic ejecta, which simulates post-merger environments, has proven to be an effective method for enhancing jet collimation \cite{2022ApJ...933L...2G, Rohoza_2024}. These dynamic ejecta contribute additional mass and magnetic fields that interact with the jet, promoting its collimation through pressure confinement and magnetic pinching effects. In our test 2D simulations, the integration of these density profiles successfully collimated the jets, although these results are beyond the scope of the current discussion and will be discussed in the forthcoming paper. Further investigations and higher-resolution simulations are currently underway to refine our understanding of jet collimation mechanisms and to enhance the precision of collimation techniques. Future work will also explore the parameter space more thoroughly, including variations in ejecta properties and magnetic field configurations, to better match observational constraints. Additionally, incorporating more sophisticated physics, such as radiation transport and neutrino cooling, may provide deeper insights into the jet formation and propagation processes. Overall, our findings demonstrate the potential of GRMHD simulations in advancing our comprehension of GRB jet dynamics. By addressing the current limitations and expanding the scope of our models, we aim to bridge the gap between theoretical predictions and observational data, thereby contributing to the broader understanding of GRBs.

\section{Acknowledgements}
This work was supported by the grant 2019/35/B/ST9/04000 from the Polish National Science Center. 
A.J. was also partially supported by the NSC grant 2023/50/A/ST9/00527. 
We gratefully acknowledge Polish high-performance computing infrastructure PLGrid (HPC Center Cyfronet AGH) for providing computer facilities and support within computational grant no. PLG/2024/017013 and PLG/2024/017347.
and Warsaw Interdisciplinary Center for Mathematical Modeling. 
We acknowledge NAOJ's support for hosting J.S. and A.J. during their visits at NAOJ. 
S.B. and M.G.D. acknowledge support from the Center for Computational Astrophysics at the National Astronomical Observatory of Japan, where the majority of the numerical computations were performed using the Cray XC50 supercomputer. 
We also thank T. Takiwaki and H. Nagakura for their useful suggestions on the density profile and Asaf Peer for the useful discussion on the jet opening angle and the 3D vs 2D analysis.

\newpage
\bibliographystyle{ws-procs961x669}  
\bibliography{references}

\vfill
\pagebreak



\end{document}